\documentclass[12pt]{iopart}

\usepackage{graphicx}
\begin{document}

\title{Belief revision in quantum decision theory: gambler's and hot hand fallacies}

\author{Riccardo Franco
\footnote[3]{To whom correspondence should be addressed riccardo.franco@polito.it}}
%\address{Dipartimento di Fisica and U.d.R. I.N.F.M., Politecnico di Torino
%C.so Duca degli Abruzzi 24, I-10129 Torino, Italia}

\date{\today}

% ----------------------------------------------------------------
\begin{abstract}
In the present article we introduce a quantum mechanism which is able to describe the creation of correlations in the evaluation of random
independent events: such correlations, known as positive and negative recency, correspond respectively to the hot hand's and to the
gambler's fallacies. Thus we propose a description of these effects in terms of qubits, which may become entangled, forming
a system which can not be described completely only in terms of its constituents.
We show that such formalism is able to describe and interpret the experimental results, thus providing a general and unifying framework
for the cognitive heuristics.
\end{abstract}

\maketitle
%
% ----------------------------------------------------------------
\section{Introduction}
This article is an attempt to provide a consistent description within the quantum formalism of the belief updating, which is a central theme in cognitive science studying how the previous information may influence the judgements (and the decisions) of the subjects.
The present work is part of a research topic which can be named \textit{Quantum Decision Theory} (for more details, see \cite{QDT}). A number of other attempts has been done to apply the formalism of quantum mechanics to domains of science different from the micro-world with applications to economics, operations research and management science, psychology and cognition, game theory, and language and artificial intelligence. For a list of references, see \cite{RF_inverse, RF_conj, RF_risk}. However, these attempts are very different and they do not provide a general way to apply the quantum formalism.

Previous quantum approaches to the problem of belief revision are Franco \cite{RF_conj} and Busemeyer \cite{busemeyer}. In particular, in \cite{busemeyer} an attempt to describe the disjunction fallacy can be considered as an approach to the concept of preference updating. However, in that article the entangled states do not describe the judgements of subjects, but the percentages of subjects who made a particular choice, and the parameters of the unitary operators do not have a clear interpretation. The works of Franco \cite{RF_inverse, RF_conj, RF_risk} focus on judgements and preferences and identify the quantum phase as a psychological parameter which can be manipulated, responsible of interference effects.

In this article  we introduce a quantum mechanism, the conditional phase shift, which is able to create strong correlations between quantum states, which may become entangled states. Such result aims to develop a method to describe human behavior within a quantum formalism, along the lines of a Quantum-like representation algorithm – QLRA \cite{khrennikov1}.
It must be stressed that the mechanism we introduce admits a clear interpretation in terms of a controlled updating of a psychological parameter, the phase, which characterizes the opinion state in our approach.

%
%
% ----------------------------------------------------------------
\section{Gambler's fallacy}
The gambler's fallacy is a widely known logical fallacy, consisting in the belief that, for random events, runs  of a particular outcome will be balanced by a tendency for the opposite outcome. In other words, it is the belief in negative autocorrelation of a non-autocorrelated random sequence of outcomes. For example, let us consider a subject repeatedly flipping a (fair) coin and guessing the outcome before it lands: if he believes in the gambler's fallacy, then after observing three heads in a row, his subjective probability of seeing another head is less than 50\%. He believes that a tail is "due," and is more likely to appear on the next flip than a head.
The gambler's fallacy is also known as \textit{negative recency}. The opposite tendency is known as \textit{positive recency} , and it is the (incorrect) belief in positive autocorrelation of a non-autocorrelated random sequence.
In the literature, we have experimental evidence for random events of positive recency, negative recency and no recency.

%Positive
Positive recency has been shown by Jarvik \cite{Jarvik}, but it becomes negative recency for longer runs.
%\cite{Lindman}????
%
In Sundali et al. \cite{Sundali} real data were considered: in 18 hours of play of a single roulette table from a large casino in Reno (Nevada), approximately half of the players exhibited gambler's fallacy (negative recency), while the other half of the players exhibited the opposite effect (positive recency).

%%%
%Negative
In \cite{Clotfelter}, data from the Maryland daily numbers game evidence a clear and consistent tendency for the amount of money bet on a particular number to fall sharply immediately after it is drawn, and then gradually to recover to its former level over the course of several months.
In  Ayton and Fisher \cite{Ayton}, a lab simulated roulette play  evidenced a gambler's fallacy tendency, which becomes stronger for longer runs: negative recency is evidenced in Experiment 1, where subjects had to gamble or forecast on colours blue or red, given a computer simulated roulette wheel.
Gambler and forecasters, though in different ways, tended to predict in the next run a different color: the longer the run of a particular color, the less likely subjects
were to predict the same color the next time. For example, after three runs of the same color, only about 46\% of subjects predicted the same color as the last outcome. The strongest gambler's fallacy in such experiment was after 5 outcomes of the same color, where 40\% of players predicted again the same color.
Moreover, it has been shown (figure 2) that the confidence expressed by subjects increases according to the runs of success (in the predictions) that they obtained in the task: we will see that this effect is connected to the hot hand's fallacy.

In Burns and Corpus \cite{Burns1} it is shown that an important factor is people's belief about the randomness of the underlying process generating the events: in some competitive contexts,  the participants  may perceive the scenario as less random, and thus they are  more likely to continue a streak. On the contrary,  with evident random scenarios the participants had a strong bias toward behavior consistent with the gambler's fallacy.
Such results suggest that one important factor influencing how people interpret sequences is how random the generating mechanism is perceived.
In the table below we illustrate the estimated probabilities by the participants in experiment of Burns and Corpus \cite{Burns1}, considering only the competitive and the random scenarios in the future case:
\\\\
\begin{tabular}{|c||c|c|}\hline
     & Competition & Random  \\\hline\hline
   Estimated probability    & $P(1)=0.55$ & $P(1)=0.45$  \\\hline
   Difference ($P(1)-0.5$) & $+0.05$ & $-0.05$  \\\hline
\end{tabular}
\\\\
The deviation form the rational judgement is equal in modulus and with different sign for the competition and the random cases.
%
%
% ----------------------------------------------------------------
\section{Hot hand's fallacy}
The hot hand's fallacy is a logical fallacy, first observed by Gilovich et al.\cite{Gilovich}, evidencing that most people associated with the game of basketball believe that a player who has just scored several times in a row is now more likely to score, because he is \textit{hot}. However, when these authors computed the sequential dependencies between successive scoring attempts of players, they found that there was no such dependency; indeed, if anything, players who have had a run of successful scoring attempts are somewhat less likely to score next time.
The hot hand fallacy can be defined  in the following general way: the hot hand is a belief in positive autocorrelation of a non-autocorrelated random sequence of outcomes like winning or losing. For example, imagine Rachel repeatedly flipping a (fair) coin and guessing the outcome before it lands. If she believes in the hot hand, then after observing three correct guesses in a row her subjective probability of guessing correctly on the next flip is higher than 50\%. Thus she believes that she is "hot" and more likely than chance to guess correctly. This particular fact has been verified in experiments of Ayton and Fisher \cite{Ayton}, where the confidence expressed by subjects increases according to the runs of success (in the predictions) that they obtained in the task.
For a review of works about hot hand's fallacy, see \cite{Bar}.
In this section, we present as an example a simple result of Gilovich, Vallone and Tversky \cite{Gilovich}, where some basketball fans considered a player who shoots 50\% from the field, and a player who shoots 70\% from the free throw line. The estimated percentage of shot after just having make a shot is respectively 61\% and 74\%. In the table below we illustrate such results:
\\\\
\begin{tabular}{|c||c|c|}\hline
  Probability of a shot  & 50\% &  70\% \\\hline\hline
  Estimated after a shot     & $P(1)=0.61$ & $P(1)=0.74$  \\\hline
  Estimated after a miss     & $P(1)=0.42$ & $P(1)=0.66$  \\\hline \hline
   Difference ($P(1)-0.5$)     & $+0.11, -0.08$ & $\pm 0.04$  \\\hline
\end{tabular}
\\\\
These results show that the deviation form the rational judgement (hot hand fallacy) is lower when the player has a higher probability of shot. We will give a quantum explanation of such fact.
%
% ----------------------------------------------------------------
\section{Relation between hot hand and gambler's fallacy}
Hot hand and gambler's fallacy can be considered as two different examples of recency effects, or of "hotness".
Subjects can believe both in the gambler's fallacy (that after three coin flips of heads tails is due) and in the hot hand (that after three wins they will be more likely to
correctly guess the next outcome of the coin toss). These biases are believed to stem from the same source, the representativeness heuristic, as discussed in \cite{Gilovich}.

Results of Sundali et al. \cite{Sundali} evidence that gambler's fallacy players are more likely to also be hot hand gamblers. These relationships, which are consistent with those previously observed in the lab \cite{Ayton}, suggest that there may be an underlying construct determining biased beliefs that further research might illuminate. In the following, we will propose a description within the quantum formalism, where we show that the recency effects may be due to the manipulation of a psychological parameter.
%
% ----------------------------------------------------------------
\section{Possible explanations}\label{explanations}
The most important explanation of the gambler's fallacy considers the operation of the representativeness heuristic \cite{Kahn1972}. It hypothesizes that people expect  the essential characteristics of a chance process to be represented not only globally in an entire sequence of random outcomes but also locally in each of its parts. Thus, despite their statistical inevitability, long runs of the same outcome lack local representativeness and are thereby not perceived as representative of the expected output of a random device. Consequently, subjects will expect runs of the same outcome to be less likely than they are: of course this is not correct, since two equiprobable outcomes are equally representative in each run.

However, the idea that perceptions of randomness are governed by representativeness has also been used to explain the exact opposite phenomenon, the hot hand fallacy.
Gilovich et al. \cite{Gilovich} supposed that judgment by representativeness leads people to reject the randomness of sequences that contain the expected number of runs because the appearance of long runs in short samples makes the sequence appear unrepresentative of randomness. Thus people tend to believe that the player will be more likely to succeed after a run of successful baskets because the player is \textit{hot}.

The explanation in terms of representativeness is quite vague in mathematical terms: it does not evidence new mathematical laws which take into account such heuristic.
The representativeness offers a convincing account of what participants do when judging random sequences but its predictive power is weak. Gigerenzer \cite{Gigerenzer} has pointed out that explaining the opposite phenomena with the same principle raises problems, yet this is what has been done with the gambler's fallacy (i.e., the streak should stop) and the hot
hand (i.e., the streak should continue).

%%%%%%%%%%%%%%%%%%%%%%%%%%%%%%%%%%%%%%%%%%%%%%%%%%%%%%%%%%%%%%%%%%%%%%%%%%5
%%%%Different perception of events --> different recency effects
%%%%%%%%%%%%%%%%%%%%%%%%%%%%%%%%%%%%%%%%%%%%%%%%%%%%%%%%%%%%%%%%%%%%%%%%%%%
%
%
Ayton and Fisher \cite{Ayton} propose a different explanation, which is confirmed by other experiments about belief revision: such effects arise because people refer to a biased concept of randomness: people have different expectations for sequences based on their experiences with different kinds of events. In fact, sequences
of outcomes reflecting human performance yield anticipations of positive recency, whereas outcomes due to inanimate chance mechanisms yield anticipations of
negative recency. This is evidenced by the fact that subjects simultaneously exhibited
both positive and negative recency (the hot hand fallacy and the gambler's fallacy) for two binary sequences with identical statistical properties.

More generally Burns and Corpus \cite{Burns1} evidenced that people are more likely to follow streaks when the mechanism generating events is nonrandom (or is believed to be nonrandom) than when the generating mechanism is random. This explanation is compatible with the one of Ayton and Fisher, since events reflecting human performance may be interpreted as non-random.

Other explanations that have been proposed are: 1) memory bias: sequence of hits are more memorable than alternating sequences, and thus their probabilities are
overestimated. Such explanation works for the hot hand fallacy, but not for the gambler's fallacy. 2) Misconception of chance.

%P(b1)=0.7
%$P(b1|c1)=P(b_0|c_0)=0.9
%P(b_1|c_0)=P(b_0|c_1)=0.1
%P(c_1)=0.75$
%
% ----------------------------------------------------------------
\section{Quantum description}
We now introduce a formalism which is able to describe both the gambler's fallacy and the hot hand's fallacy. In fact, we want to describe in a general way the belief updating of subjects about a \textit{fact} (the color of a roulette wheel or a basketball goal) or about the \textit{success} of a prediction.

Given an observable fact, we can write the dichotomous question "Will the fact happen?".
We identify with 0 the negative answer, and with 1 the positive. For example, 1 can be
the red outcome in the roulette, or a positive run of the basketball player.
In the quantum formalism, such mutually exclusive facts 0 and 1 are described by two
orthogonal vectors $|0\rangle$ and $|1\rangle$, which form a complete orthonormal
basis of a complex Hilbert space. In general, the quantum states can be in any superposition of the basis states previously presented:
\begin{equation}\label{s01}
|s\rangle=\sqrt{P(0)}e^{i\phi}|0\rangle+\sqrt{P(1)}e^{i\phi'}|1\rangle
\end{equation}
where  $\phi,\phi'$ are two phases (whose meaning will be clear further) $P(0)$, $P(1)=1$ are the estimated probabilities about the events 0 and 1
respectively, and $P(0)+P(1)=1$. Thus the vector $|s\rangle$ is called in the present article the \textit{opinion state}, since it is supposed to describe correctly the beliefs of subjects about the fact.

We now introduce a second observable, conneccted to the question is "Is my prediction  correct?". Note that such question is independent from the content of the prediction.
If the answer to such a question is positive, we denote this event with the symbol $+$, while if negative with $-$. The hot hand's fallacy happens when a $+$ situation for a previous game induces to judge that a $+$ situation is more probable than $-$ for the next game. As evidenced in \cite{Sundali} and \cite{Ayton}, players that exhibit hot hand's fallacy often also exhibit the gambler's fallacy.
We note that such second observable is somehow dual to the first observable, and connected to the concept of representativeness.
We call $P(+)$ the estimated probability that the prediction is correct, while $P(-)$ is the
estimated probability that the prediction is uncorrect.
Since $\pm$ are considered as two mutually exclusive events, we describe them as two
orthogonal vectors $|-\rangle$ and $|+\rangle$, which form a complete orthonormal
basis of a complex Hilbert space. Thus the opinion state can be also written is such basis:
\begin{equation}\label{spm}
|s\rangle=\sqrt{P(-)}e^{i\psi}|-\rangle+\sqrt{P(+)}e^{i\psi'}|+\rangle\,,
\end{equation}
where $\psi,\psi'$ are two phases (whose meaning will be clear further), $P(-)$,$P(+)$ are the estimated probabilities about the events $-$ and $+$
respectively, and $P(-)+P(+)=1$. It is important to note that the equations (\ref{s01}) and (\ref{spm}) describe the same opinion state, but they evidence different observables.

The gambler's fallacy and the hot hand fallacy can be correctly described in the quantum
formalism by using two important concepts: the interference effect and  the
conditional operations determined by the previous events.
%
%**********************************************
\subsection{The interference effect}
In a rational situation, a subject can estimate the probability of a particular outcome (for example the color of the roulette or a basket goal)  in an indirect way, by  considering the judged probabilities of correct/uncorret predictions and the conditional probabilities of the  outcome in presence of such predictions. Thus according to the classic Bayes' formula, the probability that the event will happen is $P(1)=P(-)P(1|-)+P(+)P(1|+)$, where $P(1|+)$ is the conditional probability that the event is verified, when the prediction is correct, and $P(1|-)$ is the conditional probability that the event is verified, even if the prediction is uncorrect. The Bayes' formula can be interpreted as follows: the rational subjects judge the probabilities that their predictions are correct, and from this evaluations they obtain the estimated probability of the event. If they are rational, such estimated probability is approximatively equal to the experimental data: however, in the experiments about gambler's fallacy and hot hand, this doesn't happen for the appearance of interference effects. The interpretation of such effect is the following: while estimating the probability of a given outcome, the subjects let interfere such jugdement with the question about the correctness of their prediction. Thus the judgement is influenced by psychological parameters which can modify the estimated probabilities.
%
%In a rational analysis of the color of the roulette wheel, we have a random event (0/1, red/not red) and thus $P(0)=P(1)=0.5$.  The prediction can be correct or incorrect with the same probability,  leading to $P(+)=P(-)=0.5$; moreover the conditional probabilities are $P(1|+)=P(1|-)=P(+|1)=P(-|1)=0.5$, since the fact that a prediction is correct is independent from the color of the outcome.

In a bounded-rationality situation, we suppose that the opinion state is described in the quantum framework by the opinion state $|s\rangle$ as written before. The estimated probability of the event 1 is given by $P(1)=|\langle s|1\rangle|^2$: we express  the opinion state in the basis  $|-\rangle$ and $|+\rangle$, which  is connected to the  basis of the outcomes  $|0\rangle$, $|1\rangle$ by a unitary matrix; we restrict ourselves to the following particular case:
\begin{equation}\label{transition2}
  U = \left [
  \begin{tabular}{cc}
    $\sqrt{P(0|-)}$ & $-\sqrt{P(0|+)}$ \\
    $\sqrt{P(1|-)}$ & $\sqrt{P(1|+)}$
    \end{tabular} \right ] \,
\end{equation}
This operator corresponds to the following transformation
\begin{eqnarray}\label{transformation}
  |0\rangle=\sqrt{P(0|-)}|-\rangle - \sqrt{P(0|+)} |+\rangle \\
  |1\rangle=\sqrt{P(1|-)}|-\rangle + \sqrt{P(1|+)}|+\rangle
\end{eqnarray}
The quantum formalism imposes the following constraints
$P(-|1)=P(1|-)=P(+|0)=P(0|+)$ and $P(1|+)=P(+|1)=P(0|-)=P(-|0)$, which result to be in accordance with the inverse fallacy \cite{RF_inverse}.
This means that the estimated probability that the prediction is correct when the fact is true is equal to the probability the fact is true when the prediction is correct.

Thus the opinion state in such basis is given by equation (\ref{spm}). From this formula and from the unitary matrix
(\ref{transition2}) we can express the opinion state in the basis
$|0\rangle,|1\rangle$: the complex amplitude relevant to vector $|1\rangle$ for
example is
$$\langle s|1\rangle=e^{i\psi}\sqrt{P(-)P(1|-)}+e^{i\psi'}\sqrt{P(+)P(1|+)}.$$
However, since we have to write the square  modulus of the complex amplitude, we
obtain a more general formula, which we call the interference formula:
\begin{equation}\label{interference}
P(1)=|\langle s|1\rangle|^2=P(-)P(1|-)+P(+)P(1|+)+I\,,
\end{equation}
where $I=2\sqrt{P(-)P(+)P(1|-)P(1|+)}cos(\psi-\psi')$ is the interference term, and
can be negative or positive.

We also say that if $I<0$ we have a "negative interference", since the estimated probability
about $1$ is lower than the Baynesian value. If instead $I>0$, we have a  "positive interference", since  the estimated probability about $1$ is higher than the Baynesian value.
The presence of such interference terms is responsible of the overestimation or underestimation of the probability $P(1)$.

Analogously, the classic Bayes' formula can be used to compute the probability of a successful prediction $P(+)=P(0)P(+|0)+P(1)P(+|1)$, where $P(+|1)$ is the conditional probability that the prediction is correct, when the event is verified, and $P(+|0)$ is the conditional probability that the prediction is uncorrect when the event is not verified. In the quantum formalism, such formula is modified in the following way: we consider the state (\ref{s01}), which can be written in the basis $|\pm\rangle$ through the inverse of the unitary transformation $U$ of equation (\ref{transition2}). Thus we obtain the interference formula, following the same procedure described in Franco \cite{RF_conj}:
\begin{equation}\label{interference1}
P(+)=|\langle s|+\rangle|^2=P(0)P(+|0)+P(1)P(+|1)-I'\,,
\end{equation}
where $I'=2\sqrt{P(0)P(1)P(1|-)P(1|+)}cos(\phi-\phi')$ is the interference term, and
can be  positive (overestimation) or negative (underestimation)

It is important to note that the same opinion state $|s\rangle$ may allow to produce overestimated or underestimated probabilities for $P(1)$ and $P(+)$.
%
%
%**********************************************
\subsection{Belief updating}
We now consider how the information about previous independent events can influence
(not rationally) the judgments about an event.
Let us now consider for example the color of the roulette wheel: we suppose to know
the results of the previous runs, which can be written, following the previous
formalism, with the sequence $001010110....$, where in each position 1 stands for red
and 0 for not-red (black) or goal/not goal for the basketball example. The subjects have to estimate the probability that the next outcome is $1$, knowing the result of the previous runs.

We now restrict ourselves for simplicity to a single previous run whose result is known: we want to describe the judged probability of the same outcome in  the next run. If both the first and the second game have outcomes 0, we can write the vector $|0\rangle \otimes
|0\rangle$, or simply $|00\rangle$. The symbol $\otimes$ means that we consider the
tensor product of two vectors which live in different Hilbert spaces. In other words,
for each outcome we have a different Hilbert space. In similar way, we have for the
other possible combinations the vectors $|01\rangle$, $|10\rangle$, $|11\rangle$.
Of course, we can describe the opinion state also in the basis of \textit{confidence} about the prediction $|\pm\rangle$, obtaining a superposition of states $|--\rangle$, $|-+\rangle$, $|+-\rangle$ and $|++\rangle$.

The mechanism of belief updating passes through a \textit{conditional change} of the
superposition coefficients: before knowing the result of the previous outcome, the opinion state will be
\begin{equation}\label{opinion0}
|s\rangle=|s_1\rangle |s_2\rangle\,\,,
\end{equation}
where states $|s_1\rangle$ and $|s_2\rangle$ can be expressed like in formula
(\ref{spm}), and the symbol $\otimes $ has been suppressed for simplicity. We can thus express such vector in the form
\begin{eqnarray}\label{opinion0b}
&|s\rangle=
\left[\sqrt{P(0)}e^{i\phi}|0\rangle+\sqrt{P(1)}e^{i\phi'}|1\rangle \right]
\left[\sqrt{P(-)}e^{i\psi}|-\rangle+\sqrt{P(+)}e^{i\psi'}|+\rangle \right]
\end{eqnarray}
We have expressed the first qubit in the basis of results (0,1), while the second qubit in the confidence basis ($\pm$): in fact our hypothesis is that the results of the previous runs influence the judgements about the succes of the prediction, and indirectly the judgements about the results of the next run. In particular, the information of the first run (first qubit) influences the opinion state relevant to the second run (contained in the second qubit) through a particular operator, the controlled phase shift gate (or controlled rotation or selective rotation) applied in the basis of confidence:
\begin{equation}\label{crotation}
CR=\left[
    \begin{tabular}{cccc}
    1 & 0 & 0 & 0 \\
    0 & $e^{i\chi}$ & 0 & 0 \\
    0 & 0 & 1 & 0 \\
    0 & 0 & 0 & $e^{i\xi}$
    \end{tabular}\right]\,\,.
\end{equation}
The interpretation of such operator is the following: if the first qubit is in the state 0, then the phase of the second qubit in the $\pm$ basis changes its relative phase of $\chi$. If the first qubit is in the state 1, then the phase of the second qubit in the $\pm$ basis changes its relative phase of $\xi$. Thus the probability relevant to the results of the second run changes depending on the results of the first run.

Thus, after the controlled rotation, the opinion state becomes:
\begin{eqnarray}\label{opinion1}
& CR|s\rangle=
\sqrt{P(0)}e^{i\phi}|0\rangle \left[\sqrt{P(-)}e^{i\psi}|-\rangle+\sqrt{P(+)}e^{i(\psi'+\chi)}|+\rangle\right]+\\\nonumber
&\sqrt{P(1)}e^{i\phi'}|1\rangle
\left[\sqrt{P(-)}e^{i\psi}|-\rangle+\sqrt{P(+)}e^{i(\psi'+\xi)}|+\rangle\right]
\end{eqnarray}
We note that if $\chi,\xi\neq 0$, then the opinion state can not be written as a tensor product of two opinion states relevant to two different Hilbert spaces: in this case, we have an entangled state. The interpretation of the state (\ref{opinion1}) is the following: the possible  results of the previous game influence the psychological parameter, thus producing different interference effects and modifying the estimated probabilities. In other words, the state (\ref{opinion1}) entails the following conditional judgement:
\begin{equation}\label{estim1}
\begin{tabular}{|c|c|}\hline
    Previous outcome & Estimated probability $P(1)$ for next outcome\\\hline
    1 & $P_b(1)+I(\xi)$\\\hline
    0 & $P_b(1)+I(\chi)$\\\hline
    \end{tabular}
\end{equation}
where $P_b(1)$ is the Baynesian value, and $I(\xi)$, $I(\chi)$ are the interference terms resulting from the conditional phase shift. Of course, for the estimated probability $P(0)$ we have the conditional values $P_b(0)-I(\xi)$ and $P_b(0)-I(\chi)$, which is consistent with $P(0)+P(1)=1$.
%
%
%
%
%**********************************************
\subsection{Particular cases}
If $\phi=\psi=\xi=0$ and $\chi=\psi'=\phi'=\pi$ and $P(0)=P(1)=P(\pm)=0.5$, then we have
$$
CR|s\rangle=\frac{1}{2}[|0\rangle (|-\rangle +|+\rangle)-|1\rangle (|-\rangle - |+\rangle)]
$$
Moreover, if all the conditional probabilities in transformation (\ref{transition2}) are $0.5$, we have the Hadamard transform, which leads to the state
$$
\frac{1}{\sqrt{2}}(|01\rangle-|10\rangle)
$$
which is an entangled state. This in particular evidences a situation of "complete negative recency". If the previous outcome for example is red, the judgement described by such formalism leads to predict black with certainty.

On the contrary, if  $\phi=\phi'=\psi=\psi'=\xi=0$ and $\chi=\pi$, we have
$$
\frac{1}{\sqrt{2}}(|00\rangle+|11\rangle)
$$
which is another entangled state. This in particular evidences a situation of "complete positive recency".
Of course such procedure can be done also in the basis $|\pm\rangle$: thus the entangled state $\frac{1}{\sqrt{2}}(|--\rangle+|++\rangle)$ describes a situation where a previous successful prediction leads to estimate a 1 probability of success for the following run.

In other words, the outcomes of the previous runs influence the opinion state through a conditional phase shift, which modifies the opinion state relevant to the whole sequence of runs. The resulting state is an entangled state, representing the opinion state describing the previous results and the judgements about the next run. We have shown in particular that the parameters in some special cases can lead to a maximally entangled state.
%
%
% %
%
%**********************************************
\subsection{Predictions of the model}
In this section we consider the quantitative predictions of the quantum model, and we compare it with the experimental results. First of all, we note that all the experiments of recency effects involve a description from which can be deduced the a-priori probability $P(1)$ relevant to the outcome considered. The Bayes' formula $P(1)=P(+)P(1|+)+P(-)P(1|-)$ allows to obtain  such probability from the observable $\pm$ relevant to the correctness of the prediction, and the interference effect acts as a corrective term. Thus  in our model we use the judged conditional probabilities, like $P(1|+)$ and $P(1|-)$: they give information about the judged correlation between the event 1 and the correctness  of the prediction about it. We note that such conditional probabilities in our model are supposed to follow the inverse fallacy, for which $P(+|1)=P(1|+)$; moreover, in the experimental situations they are not given, and thus they have to be estimated.

The works of Ayton et al. \cite{Ayton} and of Burns et al.\cite{Burns1} evidence that the sign of the recency effect depends on how random the outcomes of the process are perceived. If the process that generates the outcomes is explicitly random, we have negative recency. The generating process instead sems to be judged in general as non-random when it involves the human performance, as is confirmed in the experiments of Ayton and Fisher \cite{Ayton}, evidencing positive recency. However, the negative recency effect seems to be less strong when the subjects believe that the random process is generated by a precise and predictable algorithm  (forecaster case in \cite{Ayton}). Similarly, the positive recency effect is enhanced by competition contexts, and when subjects have to make judgements for future outcomes \cite{Burns1}. In other words, both negative and positive recency effects seem to be enhanced by emotive factors, such as the difficulty to perform predictions on random events ("fear effect"), or a competitive context ("hope effect"), eventually related to future events.
It is interesting to note that a maximal deviation from rational behavior  happens in the conjunction fallacy when the two outcomes are perceived as incompatible (or in antithesis) \cite{RF_conj}: the incompatibility of two situations seems to stimulate emotive factors. Analogously, in the experiments of risk and ambiguity \cite{RF_risk}, the negative interference effects appear in the comparative situation, where a vague bet follows a clear bet: the negative interference evidences a "fear effect", due to a greater difficulty to predict the possible outcome.
\\
We now show that such factors can be described in the quantum formalism by interference effects, and that the phase factor can amplify or reduce such effects. The analysis of the results of table 2 and 3 of Burns and Corpus \cite{Burns1} evidences that a competition situation leads to positive interference effect and thus to positive recency,  while  a random situation leads to negative interference terms and thus to negative recency. Moreover, the standard deviations of the estimated probabilities of streak's continuing indicate that the formal description involves mixed states containing a wide variety of states. We recall that in the quantum formalism a mixed state is a statistical mixture of different pure states associated to some probabilities: in particular, a mixed state is mathematically described by a matrix, called the density operator
\begin{equation}\label{mixed}
\widehat{\rho}=\sum_{i}P_k |s_k\rangle \langle s_k|\,.
\end{equation}
Even if the mathematical properties of the density operator are quite complicated, we observe that each pure state $|s_k\rangle$ forming the mixed state leads to a judged probability $P_k(1)$, and thus the resulting probability is $P(1)=\sum_{i}P_k P_k(1)$. Under the hypothesis that all such states differ only by relative phases in the basis $|\pm\rangle$, the resulting interference effect is
$I=\sum_{i}P_k I_k$, where $I_k$ is the interference term for the state $k$-th.
In the table below we illustrate the estimated probabilities by the participants in experiment of \cite{Burns1}, considering only the competitive and the random scenarios in the future case, which produce the strongest recency effect:
\\\\
\begin{tabular}{|c||c|c|}\hline
     & Competition & Random  \\\hline\hline
   Mean estimated probability    & $P(1)=0.55$ & $P(1)=0.45$  \\\hline
   Standard deviation    & $0.16$ & $0.13$  \\\hline
   Mean difference ($P(1)-0.5$) & $+0.05$ & $-0.05$  \\\hline
   \begin{tabular}{c}
Pure state: interference \\
$2\sqrt{P(0)P(1)P(1|+)P(1|-)}cos(\psi-\psi')=$\\
$0.5cos(\psi-\psi')$
   \end{tabular}
 & \begin{tabular}{c}
    $I=0.05$\\
$\psi-\psi'=0.93\pi$
\end{tabular} &
\begin{tabular}{c}
    $I=-0.05$\\
$\psi-\psi'=1.063\pi$
\end{tabular}
  \\\hline
\begin{tabular}{c} Mixed state (example):    \end{tabular}
 & \begin{tabular}{c}
$I_1=-0.4, P_1=0.4$\\
$I_2=0.05, P_2=0.2$\\
$I_3=0.5, P_3=0.4$\\
$\langle I \rangle=0.05$\\
\end{tabular} &
\begin{tabular}{c}
$I_1=-0.4, P_1=0.4$\\
$I_2=-0.05, P_2=0.2$\\
$I_3=0.3, P_3=0.4$\\
$\langle I \rangle=-0.05$\\
\end{tabular} \\\hline
\end{tabular}
\\\\
The deviation from the rational judgement is equal in modulus and with different sign for the competition and the random cases, but the standard deviation is higher in the competitive case: in the last row of the table above we show a possible example of mixed state which reproduces the mean estimated probability and the observed standard deviation. We note that, since the events 0/1 are completely uncertain, we consider $P(1|+)=P(1|-)=0.5$ and $P(+)=P(-)=0.5$.

We finally consider the result of Gilovich et al. \cite{Gilovich}, where some basketball fans considered a player who shoots 50\% from the field, and a player who shoots 70\% from the free throw line. The estimated shooting percentages after a shot or after a miss are shown in the table below: the hot hand effect seems to be less strong when the probability to shoot  is high.
\\\\
\begin{tabular}{|c||c|c|}\hline
  Probability of a shot  & 50\% &  70\% \\\hline\hline
  Estimated after a shot     & $P(1)=0.61$ & $P(1)=0.74$  \\\hline
  Estimated after a miss     & $P(1)=0.42$ & $P(1)=0.66$  \\\hline \hline
   Mean interference:     & $+0.11, -0.08$ & $\pm 0.04$  \\\hline
   Pure state & $I=0.5cos(\psi-\psi')$ & $I=0.29cos(\psi-\psi')$\\\hline
\end{tabular}
\\\\
In the last row of the table above we show the interference terms in the case of a pure state: in the 50\% case, we have $P(\pm)=P(1|\pm)=0.5$. In the 70\% case, we have used the arbitrary values $P(\pm)=P(1|\pm)=0.82$, since they produce the Baynesian value $P(1)=0.7$: however, they are purely indicative, in order to show a possible interference term. In both cases, the interference term is completely defined only when the phase factor $cos(\phi-\phi')$ is given, consistently with the experimental judged probabilities. Finally we note that the form of the interference term
evidences that in the case of $70\%$ the maximal interference term is always lower that the maximal interference which can be obtained in the case of $50\%$.
%
%**********************************************
\subsection{Partial trace}
The partial trace is a mathematical operation on density matrices whose physical meaning is very important. Given for example the maximally entangled state $|s\rangle=\frac{1}{2}(|01\rangle-|01\rangle)$,  we can write the density matrix $|s\rangle\langle s|$, which can be written in the standard basis $00,01,10,11$ as
\begin{equation}\label{rho_ent1}
\rho=\frac{1}{2} \left[
    \begin{tabular}{cccc}
    0 & 0 & 0 & 0 \\
    0 & 1 & -1 & 0 \\
    0 & -1 & 1 & 0 \\
    0 & 0 & 0 & 0
    \end{tabular}\right]\,\,.
\end{equation}
In particular, its elements $\rho_{ii',jj'}$ are $\langle ij|s \rangle\langle s|i'j'\rangle$,
where $i,j,i',j'=0,1$. The partial trace $Tr_1$ of the density matrix $\rho_{ii',jj'}$ over the first subsystem  is  a generalization of the trace, leading to a density matrix relevant to the second subsystem:
$$
\rho_2=Tr_1 \rho = \sum_{i=i'}\rho_{ii',jj'}
$$
In particular, for the state (\ref{rho_ent1}) we have
\begin{equation}\label{rho_ent1partial}
\rho_2=\frac{1}{2}\left[
    \begin{tabular}{cc}
    1 & 0  \\
    0 & 1
    \end{tabular}\right]\,\,.
\end{equation}
Thus we obtain a mixed state with estimated probabilities $P(1)=P(0)=0.5$, and all the interference effects are disappeared. The interpretation of partial trace is the erasure of the quantum correlation. We say that  the previous event (first qubit) has become \textit{irrelevant}, and all the interference effect have been destroyed.

This fact is connected with the intuition that gambler's fallacy and hot hand fallacy happen when subjects' judgements are \textit{emotionally} influenced by previous facts. On the contrary, when the subjects ignore previous independent results, they perform their judgements according to the rational laws: the psychological influence of the previous result on the beliefs about the next run has been erased by considering the second run as independent from the first. We can see this from equation (\ref{estim1}), by summing the estimated probabilities
$P_b(1)+I(\xi)$and $P_b(1)+I(\chi)$ relevant to the situation where the previous outcome is 1 or 0 respectively.
%
%
% ----------------------------------------------------------------
\section{Conclusions}
In the present article we have presented a purely quantum mechanism which is able to describe recency effects. In particular, we have used a two-qubit description for the autocorrelation of two non-autocorrelated events. The use of a two-qubit state for describing the opinion state evidences a situation where the subjects consider the two events as independent: the recency effect is implemented through the conditional phase shift. We evidence that such operator admits a clear interpretation in terms of a psychologic influence of the previous result. In this sense, such formalism is strictly connected to the description of the disjunction effect and of the framing effect.
Further work can be developed to consider such heuristics, as well as memory effects. In fact we note that the conditional phase shift is used as a key operation in the Grover's algorithm, which is used in the database search. Our hypothesis is that such algorithm is connected to recency effects and the availability effects.

In the description of gambler's fallacy and hot hand fallacy, the quantum correlations emerge between the opinion state of the \textit{same} subject about different facts. In this sense, we can speak of a single-subject autocorrelation. However, we note that such conditional inter-qubit effect could be used to describe also the inter-subject effects, since the quantum formalism can describe the opinion state of different subjects. This argument will be presented in a separate paper.

Finally we note that the present work and the previous study of the inverse fallacy \cite{RF_inverse} have some points in common, as evidenced by the work of Markoczy and Goldberg \cite{Markoczy}: in such work it is shown that some subjects evidence clearly an inverse fallacy, while other subjects seem to follow a pattern similar to gambler's and hot hand's fallacis. In particular, we have positive effects when human actions are considered, while negative effects for inanimate objects. This leads to make the hypothesys that subjects may update their beliefs in two distinct possible ways: a single qubit evaluation, consistent with the inverse fallacy, and a two-qubit evaluation (with a conditional phase shift), consistent with recency effects.
%
%
%%%%%%%%%%%%%%%%%%%%%%%%%%%%%%%%%%%%%%%%%%%%%%%%%%%%%%%%%%%%%%%
%
%
\footnotesize
%
%-------------------------------------------------------

% ------------------------------------------------------------------------

\section{References}
\begin{thebibliography}{99}
%
\bibitem{QDT} Site group of Quantum Decision Theory (QDT) - An exploration of quantum formalism in social and behavioral sciences, http://www.le.ac.uk/ulsm/research/qdt/

\bibitem{RF_inverse} R. Franco, The inverse fallacy and quantum formalism
http://xxx.lanl.gov/abs/0708.2972v1
%
\bibitem{RF_conj} R. Franco, The conjunction fallacy and interference effects, http://xxx.lanl.gov/pdf/0708.3948
%
\bibitem{RF_risk}R. Franco, Risk, ambiguity and quantum decision theory http://xxx.lanl.gov/pdf/0711.0886


\bibitem{busemeyer} Busemeyer, J. R., Wang, Z., \& Townsend, J. T. (2006). Quantum dynamics of
human decision making. Journal of Mathematical Psychology, 50, 220-241.

\bibitem{khrennikov1} Andrei Khrennikov,  A model of quantum-like decision-making with applications to psychology and cognitive science, http://xxx.lanl.gov/pdf/0711.1366
%%%%%%%%%%%%%%%%%%%%5
%
%Positive recency
\bibitem{Jarvik} M. E. Jarvik (1951). Probability learning and a negative recency effect
in the serial anticipation of alternative symbols. Journal of Experimental
Psychology, 41, 291-297.


%Negative recency
\bibitem{Sundali} J. Sundali, R. Croson, Biases in casino betting: The hot hand and the gambler's fallacy, Judgment and Decision Making, 1 (1), 1–12 (2006)
%
\bibitem{Clotfelter} C.T. Clotfelter, P.J. Cook, 1993. The" Gambler's Fallacy" in Lottery Play. Management Science.
%
\bibitem{Ayton} P. Ayton, I. Fisher,
The hot hand fallacy and the gambler's fallacy: Two faces of subjective randomness?
Memory \& Cognition 2004, 32 (8), 1369-1378
%
\bibitem{Burns1} B.D. Burns and B. Corpus,  Randomness and inductions from streaks:
"gambler's fallacy" versus "hot hand", Psychonomic Bulletin \& Review 11 (1), 179-184 (2004)
%%%%%%%%%%%%%%%%%%%
%HOT HAND
\bibitem{Gilovich} T. Gilovich, R. Vallone, A. Tversky, The hot hand in
basketball: On the misperception of random sequences. Cognitive
Psychology, 17, 295-314 (1985).

%%%%%%%%%Representativeness
%
\bibitem{Kahn1972} D. Kahneman, A. Tversky, Subjective probability: A judgment
of representativeness. Cognitive Psychology, 3, 430-454 (1972).
%
\bibitem{Gigerenzer} G. Gigerenzer, Surrogates for theories. In G.
Gigerenzer (Ed.), Adaptive thinking: Rationality in the
real world (pp. 289-296). Oxford, UK: Oxford University
Press (2000).


\bibitem{Gilovich1991} T. Gilovich, How we know what isn't so: The fallibility of
human reason in everyday life. New York: Free Press.

\bibitem{Tversky1989a} A. Tversky, T. Gilovich, The cold facts about the "hot
hand" in basketball. Chance: New Directions for Statistics \& Computing,
2, 16-21 (1989).

\bibitem{Bar} M. Bar-Elia, S. Avugosa, M. Raab, Psychology of Sport and Exercise 7 (2006) 525–553 Review Twenty years of ''hot hand'' research: Review and critique





%%%Belief updating
\bibitem{Markoczy} L. Mark´oczy, J. Goldberg,  Women and taxis and dangerous judgments: Content sensitive use of base-rate information, Managerial and Decision Economics, 19 (7/8) 481–493, 1998.
(Special Issue on 'Management, Organization and Human Nature')

%%%%%%%%%%%%%%%%%%%%%%%%%%%%%%%%%%%%%%%%%%%%%%%
%\bibitem{Helstrom} C. W. Helstrom, Quantum detection and estimation theory, Acd. Press,
%New York, 1976.
%
%\bibitem{Holevo} A. S. Holevo, Probabilistic and statistical aspects of quantum theory, Nauka, %Moscow, 1980; English translation: North Holland, Amsterdam, 1982.
%
%\bibitem{Nielsen} O.E. Barndorff-Nielsen, R.D. Gill, P.E. Jupp,
%On Quantum Statistical Inference, J. Roy. Statist. Soc. B 65 (2003), 775-816
%
%
\end{thebibliography}
\end{document}